\documentstyle[12pt,a4wide]{article}
\begin{document}
\newcommand{\Zzbar}
  {\mathord{\!\setlength{\unitlength}{0.9em}
  \begin{picture}(0.6,0.7)
  \thinlines
  \put(0,0){\line(1,0){0.6}}
  \put(0,0.75){\line(1,0){0.575}}
  \multiput(0,0)(0.0125,0.025){30}{\rule{0.3pt}{0.3pt}}
  \multiput(0.2,0)(0.0125,0.025){30}{\rule{0.3pt}{0.3pt}}
  \put(0,0.75){\line(0,-1){0.15}}
  \put(0.015,0.75){\line(0,-1){0.1}}
  \put(0.03,0.75){\line(0,-1){0.075}}
  \put(0.045,0.75){\line(0,-1){0.05}}
  \put(0.05,0.75){\line(0,-1){0.025}}
  \put(0.6,0){\line(0,1){0.15}}
  \put(0.585,0){\line(0,1){0.1}}
  \put(0.57,0){\line(0,1){0.075}}
  \put(0.555,0){\line(0,1){0.05}}
  \put(0.55,0){\line(0,1){0.025}}
  \end{picture}}}
\newcommand {\Cbar}
    {\mathord{\setlength{\unitlength}{1em}
     \begin{picture}(0.6,0.7)(-0.1,0)
        \put(-0.1,0){\rm C}
        \thicklines
        \put(0.2,0.05){\line(0,1){0.55}}
     \end {picture}}}

\mathchardef\Ww="3257
\mathchardef\Ss="3253
\mathchardef\Gg="3247
\mathchardef\Pp="3250
\mathchardef\Qq="3251
\mathchardef\Ll="324C
\mathchardef\Hh="3248

\begin{titlepage}
\begin{flushright} Preprint KUL-TF-92-33 \\
\end{flushright}
\vfill
\begin{center}
{\large\bf On the Generalized Miura Transformation}\\
\vskip 27.mm
{\bf Alex Deckmyn$^1$ }\\
\vskip 1cm
Instituut voor Theoretische Fysica
        \\Katholieke Universiteit Leuven
        \\Celestijnenlaan 200D
        \\B-3001 Leuven, Belgium\\[0.3cm]
\end{center}
\vfill
\begin{center}
{\bf Abstract}
\end{center}
\begin{quote}
\small
By generalizing the Miura transformation for $\Ww_N$ to other
classical $\Ww$ algebras obtained by hamiltonian reduction,
we find realisations of these algebras in
terms of relatively simple  non-abelian current algebras, e.g.
$\widehat{sl(2)}\times \widehat{u(1)}^N$, generalizing the
free field realisation of $\Ww_N$.
As an example, we present the $\widehat{sl(2)}\times
\widehat{u(1)}$ realisation of $\Ww_3^2$, which we also quantize.
By a specific example, we also show how the realisation of $\Ww_N$ with the
currents of $\Ww_{N-1}$ and a free boson can be generalized to certain
classes of ``extended'' $\Ww_N$ algebras.

\begin{flushright}
                   hepth@xxx/9209075\\
\end{flushright}
\vspace{2mm}
\vfill
\hrule width 3.cm  {\footnotesize
\noindent $^1$ Aspirant N.F.W.O., Belgium;Bitnet FGBDA23@BLEKUL11.BITNET\\}
\normalsize
\end{quote}
\end{titlepage}
\newpage

\section{Introduction}
Recently, a large class of W algebras has been shown to arise from the
hamiltonian reduction of current algebras
\cite{DrinSok,BerOog,BFOFW,FORTW,BTvD}. Specifically, different
classical\footnote{By classical algebras we will always mean Poisson
Bracket algebras.} $\Ww$ algebras can
be associated to all inequivalent $sl(2)$ subalgebras of any Lie
algebra
\cite{BTvD}. These algebras have been shown to be the symmetry algebras of
non-abelian Toda field theories or, equivalently, constrained
WZW models \cite{BFOFW}. Also, it has been shown that to classical $\Ww$
algebras and quantum $W$ algebras existing for generic central charge, a
Lie algebra can be associated with a specific $sl(2)$ embedding
\cite{BowWat,FraLin}.

To study these $\Ww$ algebras, free field realisations can be a very
powerful
tool. In fact, the well-known $W_N$ algebras were originally defined in
a realisation with $N-1$ free bosons, using the quantum Miura
transformation \cite{FatLuk}. The purpose of this paper is to
generalize this at the classical level to other $\Ww$ algebras. As it turns
out, this in general
doesn't lead directly to a free field realisation, but rather a realisation
in terms of a relatively simple non-abelian current algebra.
For example, the algebra
$\Ww_{3}^{2}$, first introduced by Polyakov \cite{Pol} and constructed
explicitely by Bershadsky \cite{Ber}, with conformal spin content
$(1,\frac{3}{2},\frac{3}{2},2)$, has a realisation with $\widehat{sl(2)}
\times \widehat{u(1)}$.

A beautiful consequence of the Miura transformation was found in
\cite{DDR}, where it was proven that $W_N$ can be realised from the
currents of $W_{N-1}$ plus an extra free boson. We generalize this kind of
reduction to other classes of $\Ww$ algebras obtained by hamiltonian
reduction.

This paper is organised as follows. In section 2, we briefly review, in an
appropriate formalism, how classical $\Ww$ algebras appear via hamiltonian
reduction. Next, in section 3, we introduce the Miura transformation
using two examples, the $\Ww_N$ series and a specific class of algebras
containing, amongst others, $\Ww_{3}^{2}$. Using this same class, we
describe in section
4 how the reduction $\Ww_N \Rightarrow  \Ww_{N-1}$ generalizes. We also
give an explicit example. In section 5 we say a few
words on the possibility of a generalized quantum Miura transformation,
and, finally, we conclude with a number of remarks.

\section{Hamiltonian reduction}

The treatment in this section is necessarily sketchy. Emphasis will be laid
on the aspects needed further along. A full treatment of classical
hamiltonian reduction in the present context was given in \cite{FORTW}.

\vspace{2 mm}

With $T_a$ being the generators of a Lie algebra $\Gg$, consider a current
\begin{equation}
J(x) = J^a(x)T_a,
\end{equation}
with
\begin{eqnarray}
J_a(x) &=& <T_a,J(x)>,\nonumber\\
g_{ab} &=& <T_a,T_b>,\nonumber\\
g_{ab}g^{bc} &=& {\delta_a}^c,\nonumber\\
J^a(x) &=& g^{ab}J_b(x),
\end{eqnarray}
where $<\;,\;>$ is the invariant bilinear form.
The current components $J_a(x)$ form a closed algebra under Poisson
Brackets (PB)
\begin{equation}
\left\{ <J(x),T_a>,<J(y),T_b>\right\}  = <[T_a,T_b],J(x)> \delta(x-y) + k
<T_a,T_b> \partial_x\delta(x-y).
\end{equation}
Now suppose we choose a subalgebra $\Ss = sl(2)$ of $\Gg$, generated by
$\left\{ M_0,M_\pm\right\}$. We can then split $\Gg$ into irreducible
representations of $\Ss$, thus obtaining a {\em graded} basis of $\Gg$
w.r.t. $M_0$:
\begin{eqnarray}
\Gg &=& \sum_{}^{} \Gg_{h_i} ,\nonumber\\
\Gg_{h_i} &=& \left\{ T \; | \; [M_0,T]=h_i T\right\} .
\end{eqnarray}
In an obvious notation, we have
\begin{equation}
\Gg \: = \: \Gg _- + \Gg _0 + \Gg _+.
\end{equation}

\vspace{2 mm}

In a first stage, we impose a number of first class constraints, the first
of which is
\begin{equation}
<J(x),M_-> =  k
\end{equation}
If $M_0$ defines an integer grading $ (h_i \in \Zzbar )$, the
constraints
\begin{equation}
\phi_{\gamma}(x) \:\equiv\: <J(x),\gamma> - k<M_+,\gamma>\: = 0\:
,\;\;
\forall \gamma \in \Gg_-
\label{allconstr}
\end{equation}
are all first class. However, if $Ad(M_0)$ also has half-integer
eigenvalues,
not all constraints with $\gamma \in \Gg_{-\frac{1}{2}}$ are first
class, since there will always be $\gamma ,\gamma '\in \Gg_{-\frac{1}{2}}$
such that
\begin{equation}
\left\{ \phi _{\gamma }(x),\phi _{\gamma '}(y)\right\}  = <J(x),M_->\delta
(x-y) + \lambda \partial _x\delta (x-y),
\end{equation}
which is not weakly zero. This is
no obstacle for the construction of $\Ww$ algebras \cite{BTvD}, but for the
Miura transformation we will make use of the gauge transformations
generated by first class constraints.
The problem of finding a complete set of first class constraints
was solved in \cite{FORTW} by introducing the notion of {\em symplectic
halving}. Using the fact that
\begin{equation}
\omega(.,.) = <[.,.],M_+>
\end{equation}
is a
symplectic form on $\Gg_{-\frac{1}{2}}$, we can apply the Darboux theorem
to split $\Gg_{-\frac{1}{2}}$ into two subspaces of equal
dimension,
\begin{equation}
\Gg_{-\frac{1}{2}} = \Pp_{-\frac{1}{2}} + \Qq_{-\frac{1}{2}},
\end{equation}
such that $\omega$ vanishes on $\Pp_{-\frac{1}{2}}$ and
$\Qq_{-\frac{1}{2}}$
seperately. This halving is by no means unique, but it is sufficiant to
choose
one particular halving. Instead of the constraints (\ref{allconstr}), we
only set
\begin{equation}
\phi_{\gamma}=0 ,\;\;
\forall \gamma \in \Gamma \equiv \Gg_{\leq -1} +
\Pp_{-\frac{1}{2}}.
\label{allconstrb}
\end{equation}
The set of constraints (\ref{allconstrb}) is now completely first class.

\vspace{2 mm}

Next, we demand these constraints to be {\em conformally invariant}. For
this, $<J(x),M_->$ will have to have conformal dimension 0 (since we
want
to constrain this component to a non-zero constant). This is not the case
when using the Sugawara tensor
\begin{equation}
L_{Sug}(x) = \frac{1}{2k} g^{ab}J_a(x)J_b(x),
\end{equation}
so we introduce the {\em improved Virasoro tensor}
\begin{equation}
L_{imp}(x) = L_{Sug}(x) + <M_0,J'(x)>.
\end{equation}

\vspace{2 mm}

After imposing the constraints, $J(x)$ is of the form
\begin{eqnarray}
J(x) &=& kM_+ + j(x) ,\nonumber\\
j(x) \in \Gamma^{\perp} &\equiv& \Gg_- + \Gg_0 +
\Qq_{\frac{1}{2}},
\label{ConstrainedCurrent}
\end{eqnarray}
where
\begin{equation}
\Qq_{\frac{1}{2}} = [M_+,\Pp_{-\frac{1}{2}}].
\end{equation}
As always, first class constraints generate gauge transformations.
\begin{equation}
J(x) \Rightarrow e^{\gamma(x)} J(x) e^{-\gamma(x)}\: + \:k
\gamma'(x),\;\;
\forall \gamma(x) \in \Gamma.
\label{gaugetransf}
\end{equation}
Using this gauge freedom we can fix $J(x)$ to a number of different gauge
choices. $\Ww$ algebras appear when choosing a so called {\em
Drinfeld-Sokolov} (DS) gauge \cite{DrinSok,BFOFW,FORTW}:
\begin{eqnarray}
J_{DS}(x)&=&kM_+ + j_{DS}(x),\nonumber\\
j_{DS}(x) &\in& g_{DS},
\end{eqnarray}
where $g_{DS}$ is a graded complement of $[M_+,\Gamma]$ in
$\Gamma^{\perp}$:
\begin{equation}
\Gamma^{\perp} = g_{DS} + [M_+,\Gamma]\;,\;\;  g_{DS}\cap\left[ M_+,\Gamma
\right] =\left\{ 0\right\} .
\end{equation}
In a graded basis $\gamma^{i}$ of $g_{DS}$,
\begin{eqnarray}
j_{DS} &=& u_i(x) \gamma ^i,\nonumber\\
\left[ M_0,\gamma^{i}\right]  &=& -h_i \gamma^i,
\label{DSgauge}
\end{eqnarray}
the components $u_i(x)$ form the algebra $\Ww_{\Gg}^{\Ss}$ under Dirac
brackets. An equivalent point
of view is to consider the $u_i(x)$ as a particular set of
{\em gauge invariant polynomials}\footnote{The fact that they are
polynomial is one of the main properties of DS gauges \cite{BFOFW,FORTW}.}
in the original current components and their derivatives.
These polynomials form the same $\Ww$ algebra under ordinary
PBs \cite{BFOFW,FORTW}.
Note that different choices of DS gauge will give other gauge fixed
polynomials, but this only corresponds to a change of basis in the $\Ww$
algebra. In the above graded basis (\ref{DSgauge}), $u_i(x)$ will have
scaling dimension
$(h_i+1)$, but it will not in general be primary. This will only be the
case in one specific DS gauge, called {\em highest weight} gauge (HW):
\begin{equation}
j_{HW}(x) = W_i(x) T^i\;,\; T^i \in ker(Ad(M_-))
\end{equation}
or
\begin{eqnarray}
<j_{HW}(x),T_i> &=& W_i(x) \;,\; T_i \in ker(Ad(M_+)),\nonumber\\
                &=& 0      \;,\; T_i \;/\!\!\!\!\!\!\in ker(Ad(M_+)),
\end{eqnarray}
where the $T_i$ is the highest weight of a spin $h_i$ representation of
$\Ss$. $W_i(x)$ is now a primary of dimension $(h_i+1)$ w.r.t. $L_{imp}$,
except for the coefficient
of $M_-$, which is basically the improved e.m. tensor $L_{imp}$. To get
$L_{imp}$ completely, one must still add the e.m tensor of the dimension 1
primaries (which correspond to $h_i=0$), as explained in
\cite{BFOFW,FORTW,BTvD}.

\section{sl(2) subalgebras}

Lie algebras generally admit a number of inequivalent $sl(2)$ subalgebras.
A particular one is the {\em principal} $sl(2)$ subalgebra, which can be
defined as $\left\{ M_{\pm},M_0\right\} $ with $M_+$ given by
\begin{equation}
M_+ = \sum_{\alpha \in \Phi _+}^{}E_{\alpha },
\end{equation}
where $\Phi _+$ is the set of positive simple roots. This particular
embedding has the additional property that the fundamental representation
of $\Gg$ is also an irreducible representation of $\Ss$. In the
case $ sl(2) \subset _p sl(N) $\footnote{$\subset _p$ denotes a
principal embedding.}, $\Ww^{\Ss}_{\Gg}$ is the well-known $\Ww_N$ algebra.
Up
to a few exceptions (in $D_n$ and $E_{6,7,8}$) all $sl(2)$ subalgebras of a
semisimple Lie algebra $\Gg$ can be found as the principal $sl(2)$ of a
regular subalgebra $\Hh$ of $\Gg$ \cite{Dynkin}. In all these cases, the
algebra $\Ww_{\Hh}^{\Ss}$ associated to the reduction scheme $sl(2)\subset
_p \Hh$ will
be a subalgebra of the final $\Ww_{\Gg}^{\Ss}$ \cite{BTvD}. For example, in
a reduction
scheme $\Ss=sl(2) \subset_p sl(n) \subset \Gg$, $\Ww_{\Gg}^{\Ss}$ will have
a $\Ww_n$ subalgebra. See \cite{BTvD} for a detailed discussion of some
reduction
schemes, and \cite{LAPP1,LAPP2} for an exhaustive list of $sl(2)$ (and
$sl(2)\times u(1)$) embeddings and their resulting $\Ww$ algebras.
\vspace{2 mm}

Throughout this paper we will use an example based on the
reduction scheme $\Ss = sl(2) \subset_p sl(N-1) \subset sl(N)$. In the
fundamental representation of $sl(N)$ and with the matrices $E_{i,i+1}$ as
positive simple roots, the constrained
current (\ref{ConstrainedCurrent}) has the form \begin{equation}
J(x) = \left( \begin{tabular}{cccccc|c}
                {*}&1&0&0&$\cdots$&0&0\\
                {*}&*&1&0&$\cdots$&&$\vdots$\\
                $\vdots$&&$\ddots$&$\ddots$&$\ddots$&&0\\
                &&&&$\ddots$&&*\\
                &&&&$\ddots$&1&$\vdots$\\
                {*}&&$\cdots$&&&*&*\\ \hline
                *&$\cdots$&*&0&$\cdots$&0&* \end{tabular} \right)\;.
\label{constraint}
\end{equation}
The topleft $(N-1)\times (N-1)$ part forms the $sl(N-1)$ algebra,
whose principal $sl(2)$ subalgebra we choose as $\Ss$, while the last
row and column form two $(N-1)$ dimensional representations of $sl(N-1)$,
and thus also of $\Ss$. Also,
the diagonal matrix $(\frac{1}{N-1},\cdots,\frac{1}{N-1},-1)$ forms a 1
dimensional representation. In the constrained current (\ref{constraint}),
there are $\left[\frac{N-1}{2}\right] $ zeros in the last column and
$\left[ \frac{N}{2}-1\right] $ in the last row.

The resulting $\Ww$ algebra has conformal spin content
$\left\{1,2,\ldots,(N-1),2\times\frac{N}{2}\right\} $, the currents of
spin $2$ to $(N-1)$ forming the subalgebra $\Ww_{N-1}$. Note that
for $N$ odd, there will be two bosonic currents with half integer conformal
spin. The simplest algebra
in this class, at $N=3$, is the algebra $\Ww_{3}^{2}$ introduced in
\cite{Pol,Ber}.

\section{The Miura transformation}
There is another useful gauge choice which is not a DS gauge:
$$J_D(x)=kM_+ + j_D(x),$$ where
\begin{equation}
j_D(x) \in \:g_{D}\: \equiv\:\Gg_0 +  \Qq_{\frac{1}{2}} +
\Qq_{-\frac{1}{2}} .
\end{equation}
In the case of
$\Ww_N$ there are no half-integer grades, and $\Gg_0$ is exactly the Cartan
subalgebra formed by the diagonal matrices (with trace zero). Hence
the name {\it diagonal gauge} (D). We will continue to use this name,
calling $g_{D}$ the {\em diagonal subspace},
but keeping in mind that $j_D(x)$ isn't purely diagonal in the general
case.
The Miura transformation now appears when comparing the D and DS gauge
\cite{BerOog}.

\subsection{Miura transformation for $sl(2)\subset _p sl(N)$}
For the present purpose, the most convenient DS gauge is not the highest
weight gauge, but what is sometimes called the {\em Wronskian} gauge
\cite{DrinSok,BerOog,BFOFW}:
\begin{equation} J_{DS}(x)=\left( \begin{array}{cccccc}
                          0&k&0&0&\cdots&0\\
                          0&0&k&0&&0\\
                          0&0&0&\ddots&\ddots&\vdots\\
                          \vdots&\vdots&&\ddots&k&0\\
                          0&0&0&&0&k\\
             u_N&u_{N-1}&u_{N-2}&\cdots&u_2&0\end{array}\right).
\end{equation}
The diagonal gauge is given by
\begin{equation}
J_D(x)=\left( \begin{array}{cccc}
                       \phi _1(x)&k&0&\cdots\\
                        0&\ddots&\ddots&0\\
                        \vdots& &\ddots&k\\
                        0&\cdots&0&\phi _N(x)\end{array}\right)
        \;,\;\sum_{i=1}^{N}\phi _i(x)=0.
\end{equation}
The fields $\phi _i$ form a closed subalgebra of $\widehat{\Gg}$, so their
Dirac brackets are identical to their PBs:
\begin{equation}
\left\{ \phi _i(x),\phi _j(y)\right\} =k\left( \delta
_{ij}-\frac{1}{N}\right) \partial _x\delta (x-y).
\end{equation}
There is a unique gauge
transformation of the form (\ref{gaugetransf}) converting $J_D$ into the
form $J_{DS}$ \cite{BFOFW,FORTW}. This would then give us the
currents $u_i$ in a polynomial realisation with the boson fields $\phi _i$.
In \cite{FORTW} a general algorithm is described to find this gauge
transformation. However, in this case one usually uses a more
subtle way to proceed \cite{BerOog}, which does
not involve finding the generator $\gamma _-$ explicitly. Consider the set
of differential
equations \begin{equation}
\left( k\partial _x-J_{GF}(x)\right) v(x)=0,
\label{diffeq}
\end{equation}
where $v(x)$ is the column matrix
$$v(x)=\left( \matrix{v_1(x)\cr \vdots\cr v_N(x)}\right) $$
If we change to another gauge using a gauge transformation
(\ref{gaugetransf}), then $v(x)$ changes according to
\begin{equation}
v(x) \rightarrow e^{\gamma (x)} v(x),
\label{v_transf}
\end{equation}
which leaves $v_1(x)$ unchanged, since $\Gamma $ contains the strictly
lower triangular matrices. If we express the set of differential equations
(\ref{diffeq}) as a differential operator $\Ll(J_{GF})$ working on
$v_1(x)$, then this differential operator is invariant under gauge
transformations:
\begin{eqnarray}
\Ll(J_{D})&=&\Ll(J_{DS}),\nonumber\\
\prod_{i=1}^{N}\left( \partial _x-\frac{1}{k}\phi _i(x)\right)
&=&\partial _x^N + \frac{1}{k}\sum_{i=0}^{N}u_i(x)(\partial _x)^{N-i},
\label{MiuraW_N}
\end{eqnarray}
where $u_0 \equiv -1$ and $u_1 \equiv 0$. The left hand side of
(\ref{MiuraW_N}) is to be read in the order
\begin{equation}
\left( \partial _x-\frac{1}{k}\phi _N(x)\right) \cdots
\left( \partial _x-\frac{1}{k}\phi _1(x)\right).
\end{equation}
Equation (\ref{MiuraW_N}) is exactly the Miura transformation for $\Ww_N$.

\subsection{Miura transformation for $sl(2) \subset_p sl(N-1) \subset
sl(N)$}
To keep a margin of clarity in the formulas, we set $k=1$\footnote{To
reinsert $k$, one simply puts a factor $1/k$ in front of every field in
the final formula.}.
The two different gauge choices are now
\begin{equation}
J_{DS}\: = \: \left( \begin{tabular}{ccccc|c}
   $\frac{H}{N-1}$& 1 & 0 & $\cdots$ & 0 & 0\\
   0 & $\ddots$ & $\ddots$ & & \vdots & \vdots \\
   \vdots & $\ddots$ & $\ddots$ & $\ddots$ & 0 & \vdots\\
   0 & $\cdots$ & 0 & $\frac{H}{N-1}$ & 1 & 0 \\
   $u_{N-1}$ & $u_{N-2}$ & $\cdots$ & $u_2$ & $\frac{H}{N-1}$
& $W_-$ \\ \hline
   $W_+$ & 0 & $\cdots$ & $\cdots$ & 0 & $-H$ \end{tabular} \right),
\end{equation}
and
\begin{equation}
J_{D}\: = \: \left( \begin{tabular}{ccccccc|c}
 $\phi _1$& 1       & 0      &$\cdots$&        &$\cdots$&0       & 0    \\
   0      &$\phi _2$& 1      & 0     &         &        & \vdots &\vdots\\
          &         &$\ddots$&$\ddots$&$\ddots$&        &        & 0    \\
   \vdots &         &$\ddots$&       &         &        &        &$J_-$ \\
          &         &        &       &$\ddots$ &$\ddots$& 0      &     0\\
   0      &         &$\cdots$&       & 0       &$\phi_{N-2}$& 1  &\vdots\\
   0      &         &$\cdots$&$\cdots$&        & 0 & $\phi _{N-1}$ & 0  \\
\hline
   0      & $\cdots$& 0      & $J_+$ & 0       &$\cdots$& 0     &$\phi _N$
\end{tabular} \right),\hspace{2 mm} \sum_{i=1}^{N}\phi _i = 0
\end{equation}
where
$J_-$ is on the$\left[ \frac{N+1}{2}\right]^{th}$ row, and
$J_+$ on the $\left[ \frac{N+1}{2}\right]^{th}$ column.

The components in the diagonal gauge again form a
subalgebra, but this time it is no longer abelian. It is easy to see that
$\left\{ \phi _i,J_{\pm}\right\} $ form the affine algebra $\widehat{sl(2)}
\; \times \; \left( \widehat{u(1)}\right) ^{N-2}$, where the
$\widehat{sl(2)}$ part is formed by $J_{\pm}$ and $J_0 \: \equiv
\frac{1}{2}\left( \phi _{\left[ N/2\right] }-\phi _N\right)$.

We again consider the set of differential equations (\ref{diffeq}) in these
two gauges. It can easily be seen that in this case, too, the
transformation (\ref{v_transf}) leaves $v_1(x)$ unchanged. So in principal
we could again consider the two differential operators on $v_1(x)$.
However, in the D-gauge this is rather cumbersome, and it turns out that
the Miura transformation in this example is most easily expressed using
Pseudo-Differential Operators\footnote{This was also done for some
classes of superconformal algebras in \cite{ItoMad}.}:
\begin{eqnarray}
\Ll(J_D) &=& \prod_{i=1}^{N-1}(\partial _x-\phi _i)
\:-\:\prod_{i=\left[
\frac{N+3}{2}\right] }^{N-1}(\partial _x-\phi _i)J_-(\partial
_x-\phi _N)^{-1}J_+\prod_{i=1}^{\left[ \frac{N-1}{2}\right] }(k\partial
_x-\phi _i),\nonumber\\
\Ll(J_{DS}) &=& -\sum_{i=0}^{N-1}u_i\left( \partial
_x-\frac{1}{N-1}H\right) ^{N-i-1} \:-\:W_-(\partial _x+H)^{-1}W_+,
\label{miuratransf}
\end{eqnarray}
where again $u_0 \equiv -1$ and $u_1 \equiv 0$.\\
The basic rules to work with these PDO's are
\begin{equation}
\left( \partial - A\right) ^{-1} = \partial^{-1} \circ \left[
1+\prod_{i=1}^{\infty }\left( A\partial \right) ^{-1}\right],
\end{equation}
and the generalized Leibniz rule
\begin{equation}
\partial^q \circ A = A\partial^{q} + \sum_{i=1}^{\infty
}\frac{q(q-1)\cdots(q-i+1)}{i!}A^{(i)}\partial ^{q-i}.
\end{equation}

For $N=3$, the result is an $\widehat{sl(2)} \times \widehat{u(1)}$
realisation of the algebra $\Ww_3^{2}$:
\begin{eqnarray}
H&=&\phi _1+\phi _2,\nonumber\\
u_2&=&J_-J_++\frac{1}{4}(\phi _1-\phi _2)^2+\frac{1}{2}(\phi
_1-\phi_2)',\nonumber\\
W_+&=&-{J_+}' +(-2\phi _1-\phi _2)J_+,\nonumber\\
W_-&=&J_-.
\end{eqnarray}
To make the underlying $\widehat{sl(2)}\times\widehat{u(1)}$ structure more
explicit, we identify $J_0=\frac{1}{2}(\phi _2-\phi _3)$ and
$U=\phi _1$. Also, we reinsert $k$.
\begin{eqnarray}
H&=&J_0+\frac{1}{2}U,\nonumber\\
T&=&u_2+\frac{3}{4k}H^2=\frac{1}{k}\left(
       J_-J_++(J_0)^2+\frac{3}{4}U^2\right) -
       \frac{1}{2} {J_0}' + \frac{3}{4}U',\nonumber\\
W_+&=&-{J_+}'-\frac{1}{k}(J_0+\frac{3}{2}U)J_+,\nonumber\\
W_-&=&J_-.
\label{ClassBer}
\end{eqnarray}

\subsection{The diagonal subspace}
As was already mentioned above, the diagonal subspace $g_D$ for both
examples is in fact
a subalgebra of $\Gg$, so we don't have to bother about Dirac Brackets.
However, this need not always be the case.
If $M_0$ defines an integral grading, there is no problem. In that case,
$g_D = \Gg_0$ is just the centralizer of $M_0$.

If $M_0$ defines a half-integral grading, however, things are not that
clear. In \cite{FORTW} so-called {\em H-compatible halvings} are
discussed.
An $sl(2)$ embedding and a particular symplectic halving are said to be
H-compatible
if a new grading operator H can be found such that\\
\begin{enumerate}
\item $Ad(H)$ has only integer eigenvalues,
\item $H-M_0$ commutes with $\Ss$,
\item dim$ker\left( Ad\left( H\right) \right) =$ dim$ ker\left( Ad\left(
M_{\pm}\right) \right)$,
\item $\Pp_{-\frac{1}{2}}+\Gg_{\leq -1}=\Gg^H_{\leq -1}$ and\\
$\Pp_{\frac{1}{2}}+\Gg_{\geq -}=\Gg^H_{\geq 1}$.
\end{enumerate}
Clearly, if an embedding and a symplectic halving are H-compatible,
$g_D$ is a subalgebra, since it is then the centraliser of H. In
\cite{FORTW} all H-compatible gradings are listed. In particular,
for any $sl(2)$ embedding of $\Gg=sl(N)$, one can always find a
H-compatible halving. In other cases, Dirac brackets must be computed.

\section{The reduction $N$ $\Rightarrow $ $N-1$}

In \cite{DDR} it was shown using the Miura transformation how the free
field realisation of $\Ww_N$ can be rewritten as a realisation with one
free boson and a $\Ww_{N-1}$ current algebra. In this section, we will
generalize their argument for the class of $\Ww$ algebras introduced in the
previous sections.

First, we will consider the case when $N$ is even (still using $k=1$).
Consider $\Ll(J_D)$ of (\ref{miuratransf}). We can then define a new set of
bosonic fields by
\begin{eqnarray}
\tilde{\phi}_i &=& \phi_i+\frac{1}{N-1} \phi _{N-1}\;,\;i=1,
\ldots ,N-2\nonumber\\
\tilde{\phi}_{N-1}&=& \phi_N + \frac{1}{N-1}\phi_{N-1},\nonumber\\
\Phi &=& \frac{1}{N-1}\phi _{N-1}.
\end{eqnarray}
The PBs of these new fields are
\begin{eqnarray}
\left\{ \tilde{\phi}_i(x),\tilde{\phi}_j(y)\right\} &=& \left( \delta
_{ij}-\frac{1}{N-1}\right)\partial _x\delta (x-y),\nonumber\\
\left\{ \tilde{\phi}_i(x),\Phi (y)\right\} &=& 0,
\end{eqnarray}
and we also have
\begin{equation}
\sum_{i=1}^{N-1}\tilde{\phi}_i=0.
\end{equation}
Writing $\Ll(J_D)$ in these new variables, we get
\begin{eqnarray}
\lefteqn{\Ll(J_D) = (\partial_x-(N-1)\Phi ) \left[
\prod_{i=1}^{N-2}(\partial _x-\tilde{\phi}_i+\Phi )
\:-\right.}\nonumber\\
&&-\left. \prod_{i=
\frac{N}{2}+1}^{N-2}(\partial_x-\tilde{\phi}_i+\Phi )J^-(\partial
_x-\tilde{\phi}_{N-1}+\Phi )^{-1}J^+\prod_{i=1}^{
\frac{N}{2}-1}(\partial _x-\tilde{\phi}_i+\Phi )\right] .
\end{eqnarray}
Using the identity
\begin{equation}
\prod_{i=a}^{b}\left(\partial_x-\tilde{\phi}_i+\Phi\right) = e^{-\Phi }
\prod_{i=a}^{b}\left(\partial_x-\tilde{\phi}_i\right) e^{\Phi },
\end{equation}
this reduces to
\begin{eqnarray}
\Ll(J_D) &=& \left(\partial_x-(N-1)\Phi \right) \:e^{-\Phi}\left[
\prod_{i=1}^{N-2}(\partial _x-\tilde{\phi}_i)\right.
\:-\nonumber\\
&&-\:\left. \prod_{i=
\frac{(N-1)+3}{2} }^{N-2}(\partial_x-\tilde{\phi}_i)J^-(\partial
_x-\tilde{\phi}_{N-1})^{-1}J^+\prod_{i=1}^{\frac{(N-1)-1}{2}
}(\partial _x-\tilde{\phi}_i)\right] e^{\Phi}\nonumber\\
&=& \left( \partial _x-(N-1)\Phi \right) e^{-\Phi }\left[ \Ll\left(
J_{D}^{N-1}\right) \right] e^{\Phi }.
\end{eqnarray}
The part in square brackets is now exactly the PDO (\ref{miuratransf}) for
extended $\Ww_{N-2}$, and can be
written in any gauge. If we put $\Ll\left( J_{DS}^{N-1}\right) $, we
thus obtain a realisation of extended $\Ww_{N-1}$ with the currents of the
extended $\Ww_{N-2}$ and the free boson $\Phi $.

\vspace{2 mm}

If $N$ is odd, the method is essentially the same, except that
we must
now extract $\phi _1$ as the free boson field instead of $\phi _{N-1}$.

So we see that in the same way that $Vir$ is the ``base'' for all $\Ww_N$
algebras (in the sence that all $\Ww_N$ algebras can be written as a
Virasoro algebra and $N-2$ commuting free bosons), $\Ww_3^{2}$ is the
base for the new class of extended $\Ww_N$ algebras introduced in the
previous sections. We might ask ourselves at this point
whether it is possible to extract the $\widehat{sl(2)}$ subalgebra too, but
looking at (\ref{ClassBer}), it does not seem possible to realise $\Ww_3^2$
with a Virasoro tensor and a $\widehat{sl(2)}$ current algebra.

\subsection*{An example: $N=4$}

As an example of this reduction, we write the algebra resulting from
the scheme $sl(2) \subset_p  sl(3) \subset sl(4)$ (which is a
particular extension of $\Ww_3$) in terms of $\Ww_3^2$ and a free boson
$\Phi $ with normalisation
\begin{equation}
\left\{ \Phi
(x),\Phi (y)\right\} = \frac{1}{12}\partial _x\delta (x-y).
\end{equation}
The currents
of $\Ww^2_3$ are written in capital letters.
\begin{eqnarray}
h&=&H+\Phi,\nonumber\\
w_+&=&W_+,\nonumber\\
w_-&=&{W_-}'-W_-(H+4\Phi ),\nonumber\\
t&=&T + \frac{1}{2}H' + 6\Phi^2 - 4\Phi',\nonumber\\
w_3&=&\frac{T'}{2} + W_+W_- +
\frac{1}{3}(H-8\Phi)T -\mbox{}\nonumber\\
&&\mbox{} - \frac{1}{12}(H-8\Phi)''
-\frac{5}{6}H H'+\frac{2}{3}(H \Phi )'+
\frac{16}{3}\Phi \Phi'-\mbox{}\nonumber\\
&&\mbox{} - \frac{7}{27}H^3 + \frac{20}{9}H^2 \Phi
-\frac{16}{9}H\Phi ^2 + \frac{128}{27}\Phi ^3.
\end{eqnarray}

Note that $w_{\pm}$ are dimension 2 primaries w.r.t. $t$, whereas $W_{\pm}$
are dimension $3/2$ primaries w.r.t. $T$.

\subsection*{Other classes of algebras}
It is now logical to ask whether there might exist other classes of
$\Ww$ algebras connected in the same way. Without giving a proof, we can
expect such behaviour to present itself for classes of algebras whose
diagonal subalgebra is $g \times \left( u(1)\right) ^{n}$, with $g$ a
semisimple Lie algebra.
It is easy to find many different examples. We present two, both based on
$sl(N)$ algebras. In the first column of the table, we give the regular
subalgebra of $sl(N)$ of which $\Ss$ is the principal $sl(2)$ subalgebra.
\vspace{2 mm}

\begin{tabular}{|c|c|c|}\hline
Regular     & Diagonal   & Spin \\
 subalgebra & subspace & content \\ \hline \hline
$ sl(N-2)$ & $sl(3)\times (u(1))^{N-3}$ &
$2,3,\ldots,(N-2),4\times 1,4\times \left( \frac{N-1}{2}\right)$ \\
\hline $ sl(N-2)\times sl(2)$ &$ (sl(2))^2 \times
(u(1))^{N-3}$ & $1,2\times 2,3,\ldots,(N-2),2\times \frac{N}{2},2\times
\left( \frac{N-2}{2}\right) $ \\ \hline
\end{tabular}
\begin{flushright}
{\em table 1. Some examples of other possible classes of extended $\Ww$
algebras.} \end{flushright}
\vspace{2 mm}

It appears that any subalgebra of $sl(N)$ of maximal rank $(N-1)$ can be
used as the diagonal subspace for a particular $\Ww$ algebra.
In particular, if $g_D$ is the Cartan subalgebra, we
get $\Ww_N$, and if $g_D$ is $sl(N)$ itself, we get the current algebra
$\widehat{sl(N)}$ (whose universal covering algebra contains all others).

\section{An example of quantization}
In the $\Ww_N$ case, the Miura transformation can be quantized directly by
interpreting all products of fields as normal ordered products. In the
general case, this is no longer possible. For instance, it is well-known
that the quantum version of the Sugawara tensor
$$L_C^{Sug}=\frac{1}{2k}g^{ab}J_aJ_b$$ is
$$L_Q^{Sug}=\frac{1}{2k+Q}g^{ab}:J_aJ_b:,$$ where $$Q g^{ab} =
\sum_{}^{}f^{ac}_df^{bd}_c.$$ So it is to be expected that the
realisations obtained in the previous sections, where the Miura form is no
longer abelian, also don't quantize trivially.
As an example of how the
coefficients can change dramatically, we present here a quantum
construction of $W_3^2$.
We choose the following normalisation for the
$\widehat{sl(2)}
\times \widehat{u(1)}$ current algebra:
\begin{eqnarray}
J_+(z)J_-(w) &=& \frac{k}{(z-w)^2}+\frac{2J_0(w)}{(z-w)}+\cdots,\nonumber\\
J_0(z)J_{\pm}(w) &=& \pm\frac{J_{\pm}(w)}{(z-w)}+\cdots,\nonumber\\
J_0(z)J_0(w) &=& \frac{k/2}{(z-w)^2}+\cdots,\nonumber\\
U(z)U(w) &=& \frac{2(k+2)/3}{(z-w)^2}+\cdots.
\label{CurrentAlgebra}
\end{eqnarray}

We now want a construction of $W_3^2$ with exactly the same terms as
the classical construction (\ref{ClassBer}), but permitting the
coefficients to change. It turns out that there is exactly one\footnote{Up
to some equivalences.} solution for which the algebra closes:
\begin{eqnarray}
H &=& J_0+\frac{1}{2}U,\nonumber\\
T &=& \frac{1}{k+2}\left( \left[ J_+J_-\right] _S+\left[
J_0J_0\right] _S
+ \frac{3}{4}\left[ UU\right] _S \right) +
\frac{1}{2}{J_0}'-\frac{3k}{4(k+2)}U',\nonumber\\
W_+ &=& -\frac{2k+1}{2}{J_+}'+\left[
J_0J_+\right] _S +\frac{3}{2}\left[ UJ_+\right] _S,\nonumber\\
W_- &=& J_-,
\label{QuantumBer}
\end{eqnarray}
where $\left[\right.$  $\left.\right] _S$ denotes {\em symmetric} normal
ordering. The formula for $H$ appears identical to the classical
case (\ref{ClassBer}), but this is a consequence of the chosen
normalisation of $U$ in (\ref{CurrentAlgebra}).

Up to a redefinition $k \rightarrow (k+1)$,
the currents (\ref{QuantumBer}) close under the same OPEs as the currents
realised in \cite{Ber}. In fact, if we realise the $\widehat{sl(2)}$
algebra using one free boson $\phi $ and a $(\beta ,\gamma )$ system
\cite{Waki},
\begin{eqnarray}
J_+&=&-\beta \gamma ^2+i \sqrt{2k+4}\gamma \partial \phi +k\partial \gamma,
\nonumber\\ J_-&=&\beta, \nonumber\\
J_0&=&\beta \gamma -\frac{i}{2}\sqrt{2k+4} \partial \phi,
\end{eqnarray}
we get exactly
the same construction as \cite{Ber}, based on two free bosons and a
$(\beta ,\gamma )$ system.

\section{Conclusions and outlook}
We have shown how the Miura transformation arises for classical
$\Ww$ algebras constructed via hamiltonian reduction. This has permitted us
to realise these algebras with a relatively simple
current algebra which we called the diagonal subspace $g_D$. In the
simplest case of $sl(N)$, the Dirac brackets of $g_D$ are identical to the
Poisson brackets, which makes it easy to identify the algebraic structure
of $g_D$.

Using this realisation, we have shown for a particular example how $\Ww$
algebras with a diagonal subalgebra of the form $\widehat{g}\times
\widehat{u(1)}^n$ can all be reduced to a certain ``base'' algebra,
realised by $\widehat{g} \times \widehat{u(1)}$, and a
number of free bosons. We expect this to be a general principle.

For the simple case $\Ww_3^2$ we have also given
the quantum version of the Miura realisation. However, in general
the quantum Miura transformation will be very hard to compute
systematically without applying the BRST methods of \cite{BerOog},
if indeed quantization is always possible without having to introduce new
terms.

In e.g. \cite{LuPop}, the reduction $W_N \Rightarrow W_{N-1}$ was also
generalized in other ways. For the classes of (quantum) algebras called
$W_N$ and $WD_N$ realisations were found not only with one smaller $W$
algebra and some free bosons, but also with a number of commuting smaller
$W$ algebras. It is natural to expect realisations of this kind to appear
in the general $\Ww_{\Gg}^{\Ss}$ case as well.

Another possibility for further investigation is the supersymmetric
version of the arguments presented in this paper. Hamiltonian reduction of
superalgebras, based on $osp(1|2)$ subalgebras of Lie superalgebras, was
discussed in \cite{LAPP2,LAPP3}. In \cite{ItoMad}, the Miura transformation
for certain classes of superconformal algebras was computed.

\vspace{1cm}
{\bf Acknowledgments}\newline
I am very grateful to LAPP in Annecy, where this work was started,
for kind hospitality, and to P. Sorba and especially E. Ragoucy for
crucial discussions. I also thank K. Thielemans and S. Schrans for many
discussions and a careful reading of this manuscript.

\end{document}